# Evidence for electromagnetic granularity in the polycrystalline iron-based superconductor LaO$_{0.89}$F$_{0.11}$FeAs

A. Yamamoto, J. Jiang, C. Tarantini, N. Craig, A.A. Polyanskii, F. Kametani, F. Hunte, J. Jaroszynski, E.E. Hellstrom, and D.C. Larbalestier
*National High Magnetic Field Laboratory, Florida State University, Tallahassee, FL 32310, USA*

R. Jin, A.S. Sefat, M.A. McGuire, B.C. Sales, D.K. Christen, and D. Mandrus
*Materials Science & Technology Division, Oak Ridge National Laboratory, Oak Ridge, TN 37831, USA*

The new rare-earth arsenate superconductors are layered, low carrier density compounds with many similarities to the high-$T_c$ cuprates. An important question is whether they also exhibit weak-coupling across randomly oriented grain-boundaries. In this work we show considerable evidence for such weak-coupling by study of the dependence of magnetization in bulk and powdered samples. Bulk sample magnetization curves show very little hysteresis while remanent magnetization shows almost no sample size dependence, even after powdering. We conclude that these samples exhibit substantial electromagnetic granularity on a scale approximating the grain size, though we cannot yet determine whether this is intrinsic or extrinsic.

Like the high temperature superconducting cuprates, superconducting oxypnictides[1,2] with the general composition REO$_{1-x}$F$_x$FeAs are layered, have a low carrier density of order $10^{21}$ cm$^{-3}$ [3] and have a parent anti-ferromagnetic (AFM) state[4-8]. Study of these compounds has been intense in the recent past and it is already established that they can superconduct when $x$ is ~0.05-0.2 and that they have transition temperatures $T_c$ above 25 K when RE = La or above 50 K when RE = Nd[9] or Sm[10]. The upper critical field $B_{c2}$ appears capable of exceeding 100 T[11], implying very short coherence lengths $\xi$ approaching the interatomic spacing, while recent muon spin rotation and NMR experiments allow deduction of a penetration depth $\lambda$ of ~0.3 μm in the low temperature limit[12,13]. In these respects also there are many similarities to the cuprates[14,15].

The similarity of basic superconducting properties suggests study of another important aspect of HTS cuprate science, namely are the grain boundaries electromagnetically granular? Here we report a study of the magnetization and microstructure of polycrystalline LaO$_{0.89}$F$_{0.11}$FeAs that shows strong electromagnetic granularity, in spite of a high residual resistance ratio (RRR) and relatively low normal state resistivity $\rho$ at $T_c$.

Polycrystalline LaO$_{0.89}$F$_{0.1}$FeAs samples of relative density ~85% were made by solid state synthesis[3]. Figure 1 shows the temperature dependence of resistivity obtained by four-probe transport measurements. The mid-point of the superconducting transition occurred at 27 K and zero resistance was observed below 22 K. The calculated resistivity values based on the nominal dimensions of 3 × 1 × 0.5 mm$^3$ are 0.153 mΩcm at 30 K and ~2.6 mΩcm at 300 K. An estimate based on the Drude formula suggests that $\rho$(300 K) should be ~1 mΩcm. Taking account of both electron and hole bands would reduce this value, suggesting that the sample is not fully connected in the normal state, even though the high RRR of ~17 suggests good metallic conductivity and reasonable normal-state connectivity.

The temperature dependent magnetic susceptibility $\chi$ in zero-field-cooling (ZFC) and field-cooling (FC) mode under an external field of 1 mT is shown in the inset of Fig. 2. The onset $T_c$ was found to be 26 K. The broad $\chi(T)$ transition may indicate either sample inhomogeneity or some electromagnetic granularity. Given a low temperature penetration depth $\lambda$ ~ 0.3 μm[12,13], we infer that broad transitions are due to flux penetrations on scales much smaller than the sample size.

Magnetization hysteresis loops and loop widths ($\Delta m$) at 5, 10, 15 and 20 K obtained by SQUID magnetometry are shown in Fig. 2 and Fig. 3, respectively. Although the sample showed a high $B_{c2}$(4.2 K) of at least 40 T[11] and a high RRR, the hysteresis loop width is always small and closes under rather low magnetic fields. Very small $M(H)$ hysteresis loops imply either very weak pinning or an imperfectly connected superconducting state. The strong paramagnetic background can be fit well by a Langevin expression consistent with almost temperature independent anti-ferromagnetism. Magneto

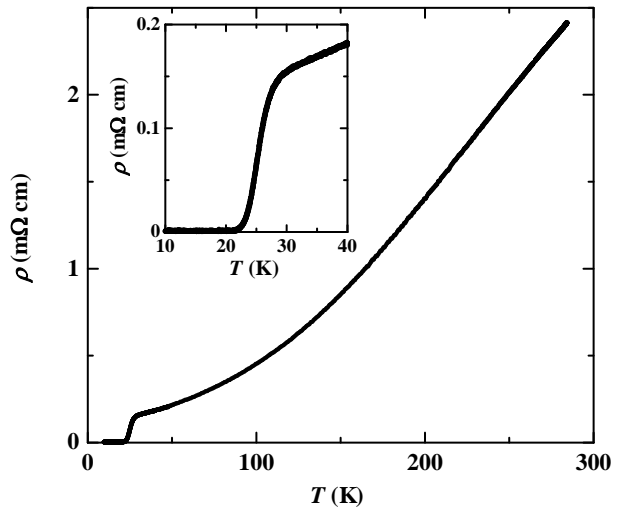

Fig. 1 Temperature dependence of resistivity for the bulk LaO$_{0.89}$F$_{0.11}$FeAs. Inset shows resistivity near $T_c$.

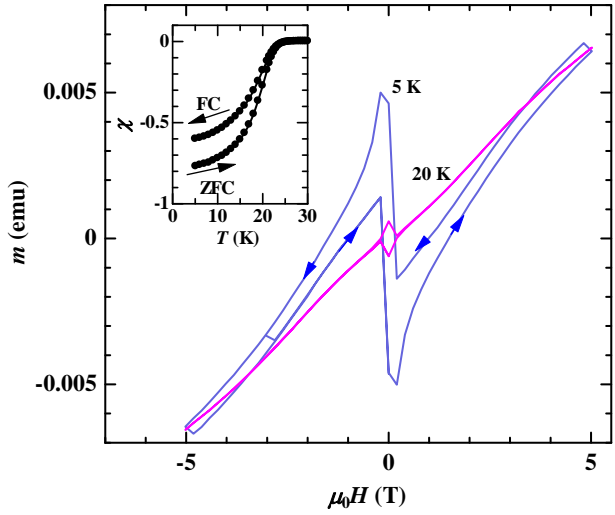

Fig. 2 (Color online) Magnetization hysteresis loops at 5 and 20 K for the bulk $LaO_{0.89}F_{0.11}FeAs$. Inset shows the temperature dependence of magnetization under zero-field-cooling (ZFC) and field-cooling (FC) conditions in an external field of 1 mT.

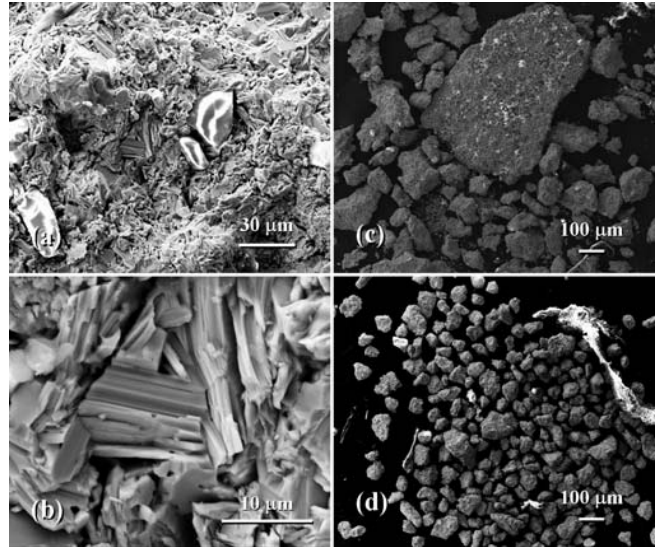

Fig. 4 Scanning electron microscope images for the $LaO_{0.89}F_{0.11}FeAs$ (a,b) bulk sample, (c) crushed pieces and (d) ground powder.

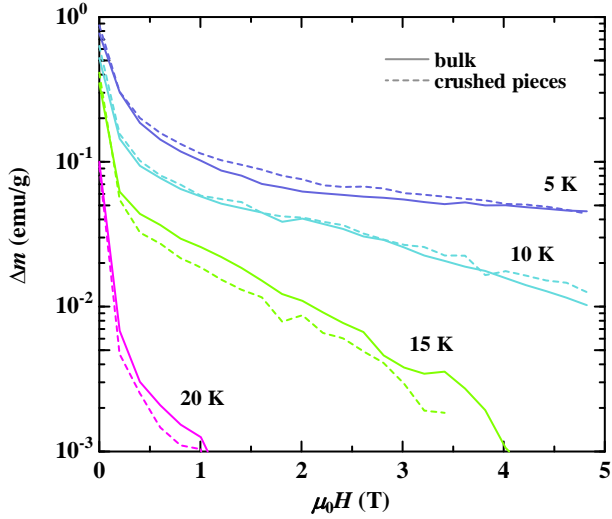

Fig. 3 (Color online) Field dependence of magnetization hysteresis loop width ($\Delta m$) at 5, 10, 15 and 20 K for the $LaO_{0.89}F_{0.11}FeAs$ bulk (solid line) and crushed pieces (dashed line). The data are normalized by sample mass.

Optical (MO) imaging was also performed to study the local magnetic structure of the bulk sample but there was almost no visible signal down to 8 K, indicating negligible whole-sample current flow, even in the most favorable case of field penetrating the whole sample and then removing the external field.

Under the assumption that the small hysteresis is due to bulk currents that can be interpreted by the Bean model, we made a more explicit test for the scale over which currents flow by exposing the whole sample to many cycles of ever increasing magnetic field $H_a$, followed by removal of the field and measurement of the remanent moment, $m_R$[16-17]. Generally flux starts to penetrate into a sample when $H_a/(1-D)$ first exceeds the lower critical field $B_{c1}$, where $D$ is the relevant demagnetizing factor. For polycrystalline samples with weak intergranular coupling, flux penetration into grain boundaries occurs at lower fields than into the grains. Therefore information about the size of current loops derived from $m_R$ (which is proportional to the product of $J_c$ and current loop size) can be extracted from the dependence of $m_R$ on the applied field $H_a$. After measurement of the whole sample, we then crushed it into tens of pieces and remeasured $m_R$, finally gently grinding the crushed pieces to powder in a mortar and remeasured $m_R$. $T_c$ did not change by crushing or powdering, nor was there any large change in the mass-normalized magnetization hysteresis $\Delta m$, as is shown in Figure 3.

Figure 4 shows scanning electron microscopy images for (a,b) the bulk sample, (c) the crushed pieces (d) and the ground powder of $LaO_{0.89}F_{0.11}FeAs$. Plate-like grains ~10 μm in size are visible in the whole sample image in Fig. 4 (b), while Figure 4(c) shows the crushed pieces in which particles of ~100 μm size and small numbers of large particles (~500 μm) were found. Figure 4(d) shows that grinding produced an average particle size of ~50 μm, which is still several times the average grain size of ~10 μm.

The remanent magnetization as a function of increasing applied field for all three sample sets is shown in Fig. 5. For the bulk sample, remanent magnetization began to increase on increasing the applied field above ~5 mT, a value consistent with the reported $B_{c1}$ of 5 mT[19]. The $m_R$ curve shows one major transition at ~20 mT and a second, minor transition at ~200 mT. This largely single-step transition implies one dominant scale of current loops, leading to the conclusion that either the sample is fully connected or that only locally circulating intra-grain supercurrents exist. For the crushed pieces, the remanent magnetization behavior was very similar to the bulk sample: the main transition peak also appeared at the same applied field of 20 mT, indicating absence of bulk-scale supercurrent in the intact bulk sample. It is even more striking that the first peak did not change after further refining the particle size to ~50 μm, indicating that currents are circulating on even smaller scales in this LaOFeAs sample.

A minor second transition was observed for the bulk and crushed pieces, however, the second peak disappeared in the ground powder sample. Its weak temperature dependence is unexplained at this time.

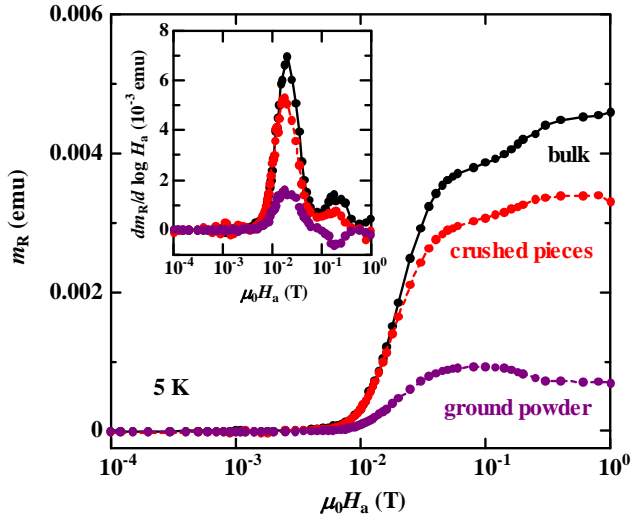

Fig. 5 (Color online) Remanent magnetization ($m_R$) as a function of applied field at 5 K for the bulk, crushed pieces and ground powder. Sample mass for the bulk, crushed pieces and ground powder are 12.0, 7.8 and 3.0 mg, respectively. Inset shows the derivatives of $m_R$.

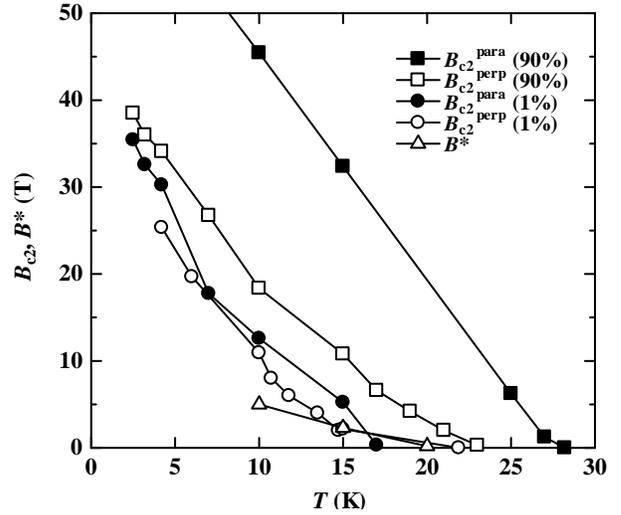

Fig. 6 Temperature dependences of upper critical field $B_{c2}$ and irreversibility field $B^*$. Filled (parallel to As-Fe plane) and open (perpendicular) squares are defined by 90% of resistive transition. Filled (parallel) and open (perpendicular) circles are defined by 1% transition.

Figure 6 shows the temperature dependences of the upper critical field $B_{c2}$[11] and the irreversibility field $B^*$, defined by $\Delta m = 10^{-2}$ emu/g, for the bulk $LaO_{0.89}F_{0.11}FeAs$ sample. Surprisingly $B^*$ was rather small compared to the very high $B_{c2}$, ~60 T at 5 K and ~20 T at 20 K. However the high $B_{c2}$ values are characteristic of the higher critical field which we presume occurs for grains with $H$ parallel to the As-Fe planes, while $B^*$ is indeed comparable to the onset of first dissipation in resistive transition curves that we associate with grains of the lower $B_{c2}$ when they are aligned perpendicular to the applied field.

In summary, magnetization hysteresis loops, microstructure and the remanent magnetization were studied as a function of sample size to test the electromagnetic granularity of a polycrystalline $LaO_{0.89}F_{0.11}FeAs$ sample. The bulk and powdered samples showed almost identical magnetization, indicating negligible bulk current in the whole sample. On the other hand, zero resistivity was observed, even in very high fields >30 T, suggesting that a true percolative supercurrent does exist across some intergranular paths. Our data are consistent with a conclusion that electromagnetic granularity occurs on a scale near that of the grain size, but that data cannot yet rule out very weak intragranular vortex pinning as the origin of the very narrow magnetization hysteresis loops. Further study is needed to decide conclusively between these alternatives and to decide whether the observed properties are intrinsic or extrinsic.

We are very grateful to Alex Gurevich and Peter Lee for discussions. Work at the NHMFL was supported by IHRP 227000-520-003597-5063 under NSF Cooperative Agreement DMR-0084173, by the State of Florida, by the DOE, by the NSF Focused Research Group on Magnesium Diboride (FRG) DMR-0514592 and by AFOSR under grant FA9550-06-1-0474. Work at ORNL was supported by the Division of Materials Science and Engineering, Office of Basic Energy Sciences under contract DE-AC05-00OR22725. AY is supported by a fellowship of the Japan Society for the Promotion of Science.